\documentclass[12pt,a4paper]{article}

\usepackage{indentfirst}
\usepackage[english]{babel}
\usepackage[T1]{fontenc}
\usepackage{textcomp} 
\usepackage[utf8]{inputenc} 
\usepackage{lmodern}
\usepackage{cite}
\usepackage{amsmath,exscale,
amssymb,latexsym,varioref,
mathrsfs,a4,textcomp,moreverb,
color,graphicx,lscape,multicol}
\usepackage{setspace}

\begin{document}
 \title{\textbf{Hydrodynamics of a quark droplet II:}\\
 {Implications of a non-zero baryon chemical potential}}
  \author{Johan J. Bjerrum-Bohr$^{1,2}$, Igor N. Mishustin$^{1,2,3}$, Thomas D{\o}ssing$^{2}$\\
 \small $^{1}$ Frankfurt Institute for Advanced Studies (FIAS),\\
 \small Goethe-University, Ruth-Moufang Str. 1, 60438 Frankfurt am Main\\
 \small $^{2}$Niels Bohr Institute, University of Copenhagen,\\
 \small  Blegdamsvej 17, 2100 K{\o}benhavn {\O}, Denmark\\
 \small $^{3}$ Kurchatov Institute, Russian Research Center,\\
 \small Akademika Kurchatova Sqr., Moscow, 123182, Russia}

\maketitle

\begin{abstract}
We present an extended version of the dynamical model for a multi-quark droplet evolution described in our proceeding paper. The model includes collective expansion of the droplet, effects of the vacuum pressure and surface tension, and now a non-zero baryon number. The hadron emission from the droplet is described following Weisskopf's statistical model. We consider evolutions of droplets with different initial temperatures and net baryon number. It is found that the introduction of a non-zero net baryon number does not change the lifetime of the droplets significantly. Only when we consider an initially very baryon-rich, low-temperature droplets is the lifetime is decreased significantly. We have, furthermore, found a convergence of both baryon chemical potential and temperature toward the values $\mu_{\rm B} \approx$ 450 MeV and $T \approx 150$ MeV. This convergence is linked to the competing emission of baryons versus mesons.
\end{abstract}

\section{Introduction}
Heavy-ion collisions studied in experiments at Brookhaven National Laboratory's Relativistic Heavy Ion Collider (RHIC) and at CERN's Large Hadron Collider (LHC) are believed to produce a new state of matter called Quark-Gluon Plasma (QGP). Several successful models describing the collision dynamics and deconfinement-hadronization transition have been proposed, but no model has so far been able to comprehend all aspects of these complicated processes. One of the most popular approaches is the hydrodynamical model first proposed by Landau \cite{Lan53} a long time ago, where the produced system is assumed to evolve as an almost perfect fluid. Different versions of this model have ever since been used to describe nuclear collisions at intermediate and high energies (see e.g. ref. \cite{Mer11} and references therein).

The most important feature of the matter produced in relativistic nuclear collisions is its strong collective expansion which is well established experimentally.(See e.g. \cite{Mul12} and references therein).

In refs. \cite{Mis99b, Mis09}, the qualitative arguments have been presented that a 1st order phase transition in a rapidly expanding system should precede via overcooling and subsequent fragmentation of the plasma phase into quark droplets surrounded by a low-density hadron gas. Such scenario was further studied in refs. \cite{Mis99a,Cse95, Sca01,Ran04,Mis08}. Recently, in a series of publications \cite{Nah13,Her13a,Her13b} this qualitative picture was confined within a chiral fluid dynamics model including fluctuations and dissipative terms. In our preceding publication \cite{Bje12} (henceforth denoted as (I)), we have developed a simple model which describes the evolution of a quark droplet taking into account the hadron emission from the surface. A similar model was considered in earlier investigations \cite{Bir83,Bar90}, where a schematic version of a hydrodynamical model was first introduced.

The key features of the approach formulated in (I) include:
\begin{enumerate}
\item The model adopts the phenomenological description of quark-droplet using the MIT bag model.
\item The quark and antiquarks inside the droplet are described by hydrodynamics as a perfect fluid. 
\item A surface energy associated with the gradients of the mean fields (condensates) in the transition region between inside and outside of the droplet is included in the total energy balance. 
\item The hadronization of the droplet is assumed to proceed via emission of hadrons through the surface described by Weisskopf's statistical model. 
\end{enumerate}\ \\
In (I) we have shown, that if hadron emission is disregarded, the droplet behavior is described by anharmonic oscillations, which can be approximated by a super-ellipse parametrization of the phase-space trajectories. In the small-amplitude limit, we found that the period of oscillations is proportional to the droplet radius. When hadron emission is included, the motion is changed to damped oscillations characterized by the decay time, which in the first approximation is proportional to the droplet radius too. For droplets with initial radii 1.5-2 fm the decay time lies in the interval 9-16 fm/$c$, depending on parameters of the initial state.

Our calculations have shown that pions and kaons are by far the most abundant emitted particles, since the emission of heavier species is suppressed by their mass. Nevertheless, the heavy particle spieces contribute significantly to the energy loss as the emitted pions carry away only about 50 \% of the total energy. We also found damped oscillations of the droplet due to the initial expansion.

In (I), we have only considered baryon-free droplets made of quarks and antiquarks with zero chemical potentials. This is a reasonable assumption for plasma produced at RHIC and LHC energies \cite{And11} but at lower collision energies the effects due to the net-baryon dynamics become more important. So, in the present paper, we present new results obtained for non-zero baryon chemical potential. This study could lead to better understanding of such interesting phenomena as cold stranglet formation and flavor distillation \cite{Bar90,Gre87,Gre88}. It turns out that for a non-zero baryon chemical potential the model has to be completely reformulated, since a new numerical method to solve the set of equations is required. 
We did not include the strangeness chemical potential in the present study, since the additional dimension would make calculations even more challenging.

The present paper is organized as follows: In Sect. \ref{sec:Theory}, we will briefly summarize the essential formulas of the model including the hadron emission rates from the surface of the droplet. In Sect. \ref{sec:Numerics}, we will schematically explain how the numerical simulations are carried out. Then in Sect. \ref{sec:Results} we present results for the droplet evolution for different initial values of temperature and baryon chemical potential. Conclusion and outlook are given in Sect. \ref{sec:Conclusion}.

\section{Modeling evolution of a quark droplet} \label{sec:Theory}
\subsection{The global dynamics}
Below we describe finite droplets of quark-antiquark plasma following the MIT bag model \cite{Cho74,DeG75}, i.e. we assume that massless quarks and antiquarks are confined in a spherical cavity with radius $R$ and bag constant $B$.

The evolution of the droplet is described by the hydrodynamic equations for a perfect fluid characterized by a spherically-symmetric velocity field,
\begin{equation}
v(r) = \dot{R} \frac{r}{R} \label{LinVel}
\end{equation}
isotropic pressure $P$ and proper energy density $\epsilon$. Both $\epsilon$ and $P$ are assumed to be homogenous and depend on temperature $T$ and chemical potential of quarks $\mu$ only.

We assume also that the droplets are surrounded by a dilute hadron gas or even by the physical vacuum. Therefore, to account for the gradients of the macroscopic quantities, a surface energy is added to the total energy of the droplet,
\begin{equation}
E_{\rm surface}(R,\dot{R}) = \frac{4\pi R^2 \sigma}{\sqrt{1-\dot{R}^2}},
\end{equation} 
where $\dot{R}$ is the surface velocity and $\sigma$ is the surface tension coefficient. This expression can be interpreted as the energy of a relativistic particle of mass $4\pi R^2\sigma$ moving with velocity $\dot{R}$.

In general the bag constant $B$ and surface tension $\sigma$ should depend on $T$ and $\mu$. In particular, the $T$ dependence of $B$ at small $\mu$ can be extracted from the lattice calculations. This information can be in principle obtained from the interaction measure, $\epsilon-3P$, see e.g. refs. \cite{Dum12,Beg10}. However, there is still a significant uncertainty in the interpretation of lattice results. There were attempts to replace $B$ by an additional scalar field (dilaton), but resulting models have other parameters which are difficult to fix at present, see e.g. \cite{Sas12}. The behavior of $\sigma$ as function of $T$ and $\mu$ is even less known, see discussion of this problem in (I). By this reason in the present study we use these parameters the constant values $B$=200 MeV/fm$^3$ and $\sigma$=50 MeV/fm$^2$, which lie somewhere in the middle of the uncertainty interval.

The global dynamics of the droplet is derived from averaging the hydrodynamic equations of energy-, entropy- and net baryon number-conservation over the spatial coordinates, similar to the approach of Refs. \cite{Bir83,Bar90}.

With the surface contribution the equations for the global dynamics of the droplet (without particle emission) can be written as conservation laws for total energy, total entropy and the net baryon number (see more details in (I)):
\begin{align}
& E(R,\dot{R},T,\mu) = \frac{4\pi}{3} R^3 \left[ (\epsilon + P) \langle \gamma^2 \rangle- P + B \right] + E_{\rm surface}(R,\dot{R}), \label{Energy} \\
& S(R,\dot{R},T,\mu) = \frac{4\pi}{3} R^3 s \langle \gamma \rangle,\label{Entropy} \\
& N_{\rm B}(R,\dot{R},T,\mu) = \frac{4\pi}{3} R^3 n_{\rm B} \langle \gamma \rangle. \label{BaryonChemPot}
\end{align}
Here $s$ is the entropy density and $n_{\rm B}$ is net baryon density. The explicit expressions for the spatial averages of $\gamma$ and $\gamma^2$ factors for the linear velocity profile of Eq. (\ref{LinVel}) are given by:
\begin{align}
\langle \gamma \rangle &= \frac{3}{2} \frac{1}{\dot{R}^3}\left(\arcsin(\dot{R})-\dot{R}\sqrt{1-\dot{R}^2}\right),\label{gamma}\\
\langle \gamma^2 \rangle &=  \frac{3}{\dot{R}^3}\left(\text{arctanh}(\dot{R})-\dot{R}\right). \label{gammaSQR}
\end{align}
The energy density, pressure, entropy density and baryon density of the plasma are expressed as\footnote{Note that the bag pressure is already included in Eq. (\ref{Energy}) } 
\begin{align}
\epsilon(T,\mu)&= \frac{7 \pi^2}{120} \nu_q T^4 + \frac{\nu_q}{4} T^2\mu^2 + \frac{\nu_q}{8\pi^2} \mu^4 , \qquad P(T,\mu)=\frac{1}{3}\epsilon(T,\mu), \label{ensdens}\\ 
 s(T,\mu)&=\frac{7\pi^2}{90} \nu_q T^3 + \frac{\nu_q}{2}T\mu^2,\label{entdens}\\ 
 n_{\rm B}(T,\mu) &= \frac{\nu_q}{18} T^2\mu + \frac{\nu_q}{18}\frac{\mu^3}{\pi^2} \label{numdens},
\end{align}
where $\nu_{q}$ is the degeneracy factor of quarks (antiquarks). Since we include: up, down and strange quarks, $\nu_q = 18$ in our calculations. 

\subsection{Formulation of hadron emission}
The idea behind the emission of hadrons through the surface of the droplet is that quarks close to the surface can combine into a colorless hadron, which may leave the droplet carrying away the energy $\omega_h = \sqrt{m_h^2 + p^2}$, where $m_h$ is the mass of the hadron. The ability to emit hadrons is determined by the thermal excitation energy $E^\star(T,R,\dot{R})= E(T,R,\dot{R})-E(0,R,\dot{R})$ and the droplet is assumed to be in thermal equilibrium before and after each emission. We is describe the process following Weisskopf's statistical model \cite{Wei37}; a similar approach has been used also in ref. \cite{Bar90}. According to this approach, the double-differential emission rate is given by the ratio of the statistical weights before and after the emission, $\Omega_{\rm after}/\Omega_{\rm before}$:
\begin{equation}
\frac{d^2N_h}{dp \ dt} = \frac{1}{2} \frac{\nu_h}{4 \pi^2} p^2 \mathcal{A}_{\rm geom} \exp[\Delta S],
\end{equation}
where $p$ is momentum of the emitted hadron, $\nu_h$ is its spin-isospin degeneracy factor, $\mathcal{A}_{\rm geom}=4\pi R^2$ is the total surface area of the droplet, and $S(E^\star,R)$ is the entropy of the droplet at the thermal excitation energy $E^\star$ and volume $V=\frac{4}{3}\pi R^3$. Note that we have introduced a factor of 1/2, since the particle emission is only possible for polar angles $\theta < \frac{\pi}{2}$.
The change in the entropy due to emission of a hadron $h$ can be found from the 2nd law of thermodynamics:
\begin{equation}
\Delta S_h = \frac{1}{T} \left(\Delta E^\star_h + P \Delta V_h - \mu_h \Delta N_h \right),
\end{equation}
where, $\Delta E_h^\star = \omega_h$ is the change of the 
excitation energy, $\Delta N_h = -1$ is the change of the number of hadrons of type $h$, and $\Delta V_h$ is the change of the volume due to the particle $h$ emission (neglected in the present calculation).
Now we can calculate the baryon number and energy emission rates
\begin{align}
\left[\frac{dN_{\rm B}}{dt}\right]_{\rm emit} &= \sum_h \frac{1}{2}\frac{\nu_h}{2\pi^2}  b_h \mathcal{A}_{\rm geom} \int^\infty_0 p^2 e^{-\frac{\sqrt{m_h^2+p^2}-b_h\mu_{\rm B}}{T}} \ dp, \label{averageNloss}\\
\left[\frac{dE}{dt}\right]_{\rm emit} &= \sum_h \frac{1}{2}\frac{\nu_h}{2\pi^2} \mathcal{A}_{\rm geom} \int^\infty_0 p^2 \sqrt{m_h^2 + p^2} e^{-\frac{\sqrt{m_h^2+p^2}-b_h\mu_{\rm B}}{T}} \ dp, \label{averageEloss}
\end{align}
where $b_h$ is the baryon charge of the particle $h$. In our calculations, we include emission of mesons ($\pi, K, \rho, \omega$), baryons $(N, \Delta)$ and hyperons ($Y=\Lambda, \Sigma$). Obviously, the particle emission leads to changes of the total energy, entropy and baryon number of the droplet. Therefore, we should now solve the following set of equations
\begin{align}
\frac{dE}{dt} &= \frac{d}{dt} \left[ V \{ (\epsilon+P)\langle \gamma^2 \rangle - P + B\} + \frac{4\pi R^2 \sigma}{\sqrt{1-\dot{R}^2}}\right] = - \left[\frac{dE}{dt}\right]_{\rm emit},\label{Eloss}\\ 
\frac{dS}{dt} &= \frac{d}{dt}\left[Vs \langle \gamma \rangle \right] = - \frac{1}{T} \left( \left[\frac{dE}{dt}\right]_{\rm emit} -\mu_{\rm B} \left[\frac{dN_{\rm B}}{dt}\right]_{\rm emit}\right), \label{Sloss}\\
\frac{dN_{\rm B}}{dt} &= \frac{d}{dt}\left[Vn_{\rm B} \langle \gamma \rangle \right] = - \left[\frac{dN_{\rm B}}{dt}\right]_{\rm emit}, \label{Nloss}
\end{align}
where $\mu_{\rm B}= 3 \mu$ is the baryon chemical potential.
These equations together with Eq. (\ref{averageEloss}) and (\ref{averageNloss}) should be solved for functions $R(t)$, $T(t)$ and $\mu_{\rm B}(T)$. A more detailed description of the model is given in (I).

Finally, it is interesting to estimate the recoil pressure on the droplet's
surface due to the hadron emission, see also \cite{Bar90}. It can be calculated as the average radial momentum $p \cos{\theta}$ transferred to the droplet by emitted particles per unit time and unit area. The corresponding expression can be obtained from Eq. (\ref{averageEloss}) by replacing $\omega_h$ by $p/(2\cal{A}_{\rm geom})$. Here the factor 1/2 comes from the averaging over azimuthal angles in the interval $0<\theta<\pi/2$. The estimated upper bound of the recoil pressure for droplet radius $R$=2 fm is 30 MeV/fm$^3$, that is much smaller than the vacuum pressure (200 MeV/fm$^3$) and Laplace pressure (50 MeV/fm$^3$). Therefore, we neglect this contribution in further calculations.

\section{Numerical simulations of droplet dynamics}\label{sec:Numerics}
For the numerical solution of these equations one should specify initial conditions and develop a numerical algorithm to determine the time evolution of the droplet parameters. To solve Eq. (\ref{Eloss}), (\ref{Sloss}) and (\ref{Nloss}), we have created a numerical method which at each time step ($n \cdot \Delta t$) determines the variables at the next time step ($(n+1) \cdot \Delta t$).

At the first time step we set the initial parameters for radius $R$, velocity $\dot{R}$, temperature $T$ and net baryon number $N_{\rm B}$. Using these parameters, we can from Eqs. (\ref{ensdens})-(\ref{numdens}) determine the initial baryon chemical potential $\mu_{\rm B}$, total entropy $S$, total energy $E$ etc.

To determine the parameters at the next time step we use the following procedure:
\begin{enumerate}
\item{Knowing $R$, $\dot{R}$, $T$ and $\mu_{\rm B}$, we can determine $dE/dt$ and $dN_{\rm B}/dt$ from Eq. (\ref{averageEloss}) and (\ref{averageNloss}) and determine $dS/dt$ from Eq. (\ref{Sloss}).}
\item{The acceleration $\ddot{R}$ can, now, be calculated by rewriting the total energy as a function of $R$, $\dot{R}$, $S$ and $N_{\rm B}$: 
\begin{equation}
\frac{dE}{dt} = \frac{\partial E}{\partial R} \dot{R}  + \frac{\partial E}{\partial \dot{R}}\ddot{R} + \frac{\partial E}{\partial S} \frac{dS}{dt} + \frac{\partial E}{\partial N_{\rm B}} \frac{dN_{\rm B}}{dt},\label{Acceleration}
\end{equation}
and finding the partial derivatives of the total energy: $\frac{\partial E}{\partial R}$, $\frac{\partial E}{\partial \dot{R}}$, $\frac{\partial E}{\partial S}$, $\frac{\partial E}{\partial N_{\rm B}}$.}
\item{From $S$, $dS/dt$, $N_{\rm B}$ and $dN_{\rm B}/dt$ we calculate $S$ and $N_{\rm B}$ at the next time step.}
\item{Then we use the acceleration to find $R$ and $\dot{R}$ using Runge-Kutta method and finally we use $R$ $\dot{R}$, $S$ and $N_{\rm B}$ to determine $T$ and $\mu_{\rm B}$ at the next time step through Eq. (\ref{Entropy}) and (\ref{BaryonChemPot}). Thereafter the calculations is repeated for the next time step $(n+1)$.} 
\end{enumerate}

It turned out that a resulting expression for the acceleration $\ddot{R}$ contains a singularity for $\dot{R}=0$. We believe this is a consequence of our simplified treatment of the dynamics. Obviously, such a term should vanish in the exact solution. Fortunately, this pole has a small coefficient and can be neglected with reasonable accuracy.

\subsection{Choice of initial parameters}
To study the dynamics of the quark droplet, we have to specify the model parameters bag pressure and surface tension and initial conditions, i.e. initial radius, collective expansion velocity, initial temperature of the quark plasma and the net baryon number of the droplet. In this paper, we will only study close-to-equilibrium scenarios of the droplet evolution, i.e. a non-expanding droplet ($\dot{R}=0$) where initially the thermal pressure is balanced by the vacuum pressure (bag constant) outside the droplet and the Laplace pressure from the surface. This balance can be described by the equation for vanishing net pressure acting on the surface
\begin{equation}
P_{\rm net}(R,T,\mu_{\rm B})= P(T,\mu_{\rm B}) - B - \frac{2\sigma}{R} = 0, \label{PressureBalance}
\end{equation}
Implication of the initial expansion was studied in (I). By studying close-to-equilibrium scenarios, we will focus on the new features of the dynamics associated with a finite baryon chemical potential. As in (I), we take an initial droplet radius $R \approx $2 fm, no initial velocity $\dot{R}_0\approx 0$.

Below we study 5 cases of the evolution for different sets of initial temperature and baryon chemical potential number. We choose these sets such that at initial stage, all of them have the same bulk energy density of the quark-antiquark plasma. Since the vacuum energy and surface energy depend only on the radius for $\dot{R}=0$, the initial total energy, $E_{\rm t}$, of the droplet will be the same too. In order to compare with our previous results from (I), we choose a droplet with initial energy density $\epsilon = 0.75$ GeV/fm$^3$. This means a total energy $E=$34.4 GeV, with a quark-antiquark plasma energy contribution $E_{\rm p}=$ 25.1 GeV, a vacuum energy contribution $E_{\rm vac}=$ 6.7 GeV and finally a surface energy contribution $E_{\rm sur}=$ 2.5 GeV. We will denote each of the cases with a number from 1 to 5. The initial choice of temperatures and net baryon numbers for different cases are shown in table (\ref{cases}). The first case corresponds to the zero baryon chemical potential, close-to-equilibrium, studied in our previous paper (I). We stop the calculations, when the radius of the droplet reaches 0.8 fm, which is close to the proton radius. At this point, we can no longer justify our microscopic considerations. 

\begin{table}[htdp]
\begin{center}
\begin{tabular}{|c|c|c|c|c|}\hline
 & \multicolumn{2}{|c|}{Chosen parameters} & \multicolumn{2}{|c|}{Corresponding parameters} \\ \hline
Case & $N_{\rm B}$ & $T$ [MeV]   & $n_{\rm B}$ [fm$^{-3}$] & $\mu_{\rm B}$ [MeV] \\ \hline
1 & 0   & 153.4 &  0  & 0          \\ \hline 
2 & 7   & 149.9 &  0.2 & 208.7 \\ \hline 
3 & 13 & 140.0 &  0.4 & 413.4 \\ \hline 
4 & 20 & 109.6 &  0.6 & 747.7 \\ \hline 
5 & 23 & 84.3   &  0.7 & 932.1 \\ \hline
\end{tabular}
\caption{\small The initial net baryon number and temperature together with the corresponding initial baryon density and chemical potential.}\label{cases}
\end{center}
\end{table}

\section{Results and discussions}\label{sec:Results}
\begin{figure}
\centering
\includegraphics[width=\linewidth]{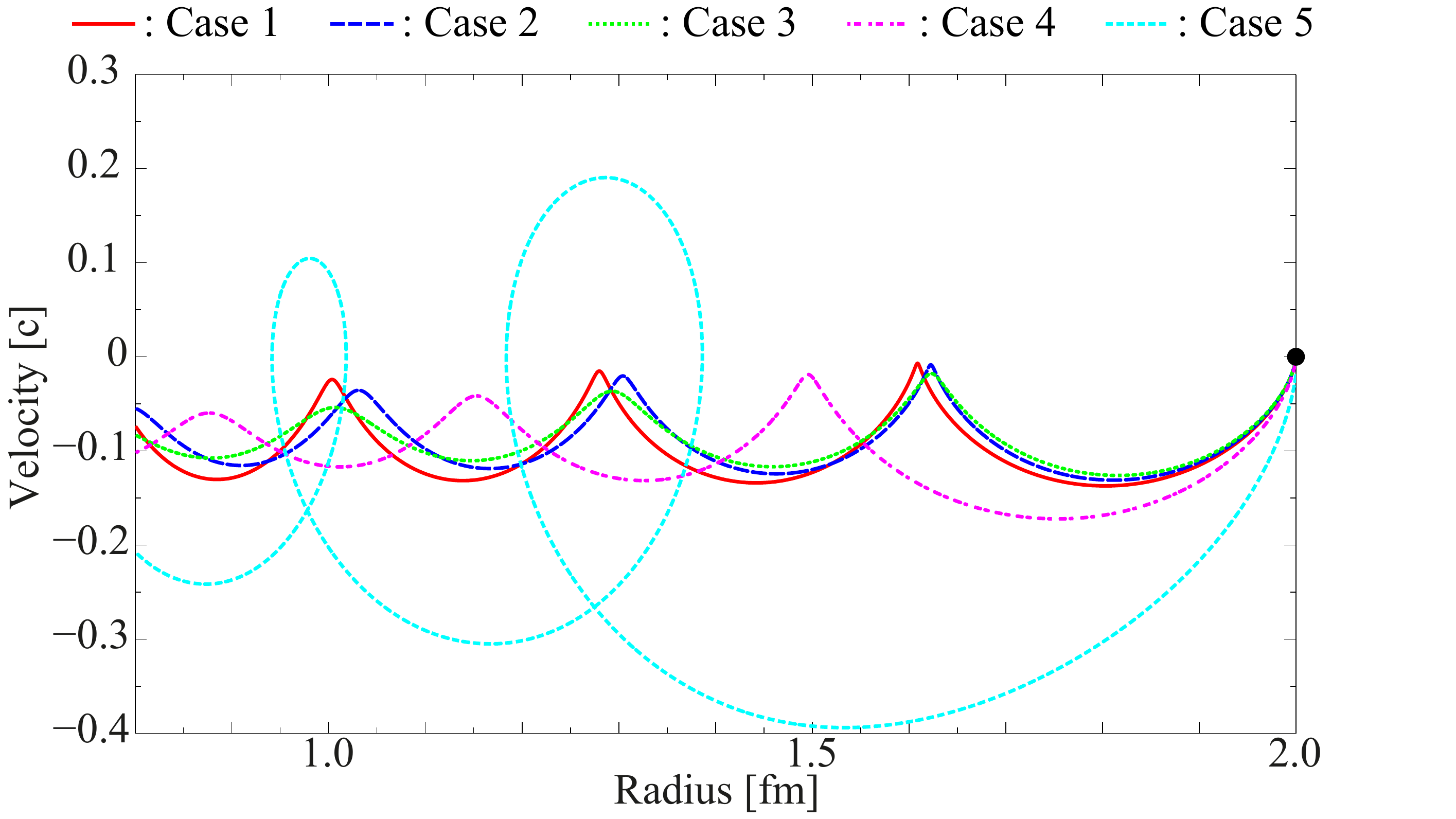}
\caption{\small (color online) Trajectories of the quark droplets in the $R - \dot{R}$ plane are shown for four different initial conditions (as indicated in the figure). The dots indicate the initial point of the droplet evolution. A cut-off at $R = 0.8 \rm \ fm$ is applied.}\label{plot:radiusVSvelocity}
\end{figure}

\begin{figure}
\centering \includegraphics[width=\linewidth]{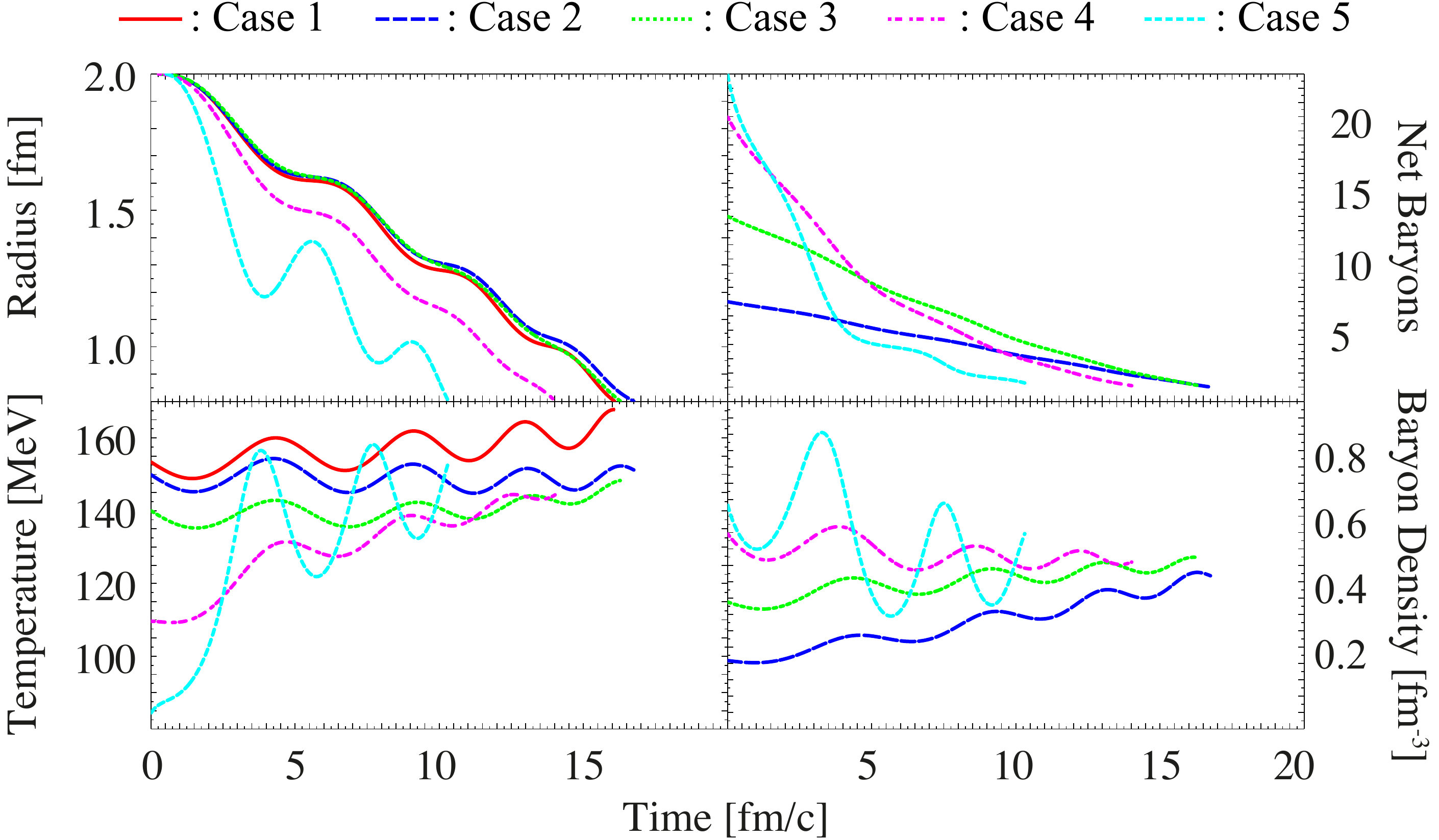}
\caption{\small (color online) The droplet radius, the temperature of the quark plasma, the net baryon number inside the droplet and the net baryon density as functions of time for five different initial conditions from Table \ref{cases}. The calculations are stopped when the radius reaches 0.8 fm.}\label{plot:evolution}
\end{figure}

In this section, we present results for 5 cases of droplet dynamics with initial parameters chosen in accordance with Table \ref{cases} In Fig. \ref{plot:radiusVSvelocity} we show phase trajectories, i.e. the surface velocity as a function of radius of the droplet for these 5 cases. The dynamics shows an oscillatory behaviour superimposed with the shrinkage of the droplet due to the hadron emission, similar to the picture found in (I). In cases 1 to 4, the droplet does not acquire any outward speed at all. In case 5, on the other hand, the oscillations of the droplet are much stronger, leading to inwards surface velocities up to $0.4c$, similar to the out-of-equilibrium cases considered in (I). This case shows that although the pressures acting on the droplet is in perfect balance initially, the evaporation of hadrons may be so strong that the dynamics is shifted further away from the equilibrium compared to the other cases. This is shown more directly in the next figure, Fig. \ref{plot:evolution}. Here, the droplet radius, the temperature of the quark plasma, the net baryon number and the net baryon density are shown as functions of time for the same 5 cases. Again for the radius as a function of time, in cases 1 to 4, we find a superposition of small oscillations with a gradual shrinkage, looking almost like a linear drop, with rates of about $0.075c$ for case 1 to 3 and $0.086c$ for case 4. However, for case 5, we find strong damped oscillations. From the plot of net baryons as function of time, we see initially a very strong emission of baryons in both case 4 and especially in case 5, where almost 12 and 20 baryons respectively are emitted during the first 5 fm/$c$. This shows that the large initial baryon chemical potential strongly enhances the net baryon emission during the initial stages of the droplet evolution. This gives a strong offset between different contributions to the pressure leading to a more violent dynamics seen in both case 4 and especially case 5.

\begin{figure}
\centering
\includegraphics[width=\linewidth]{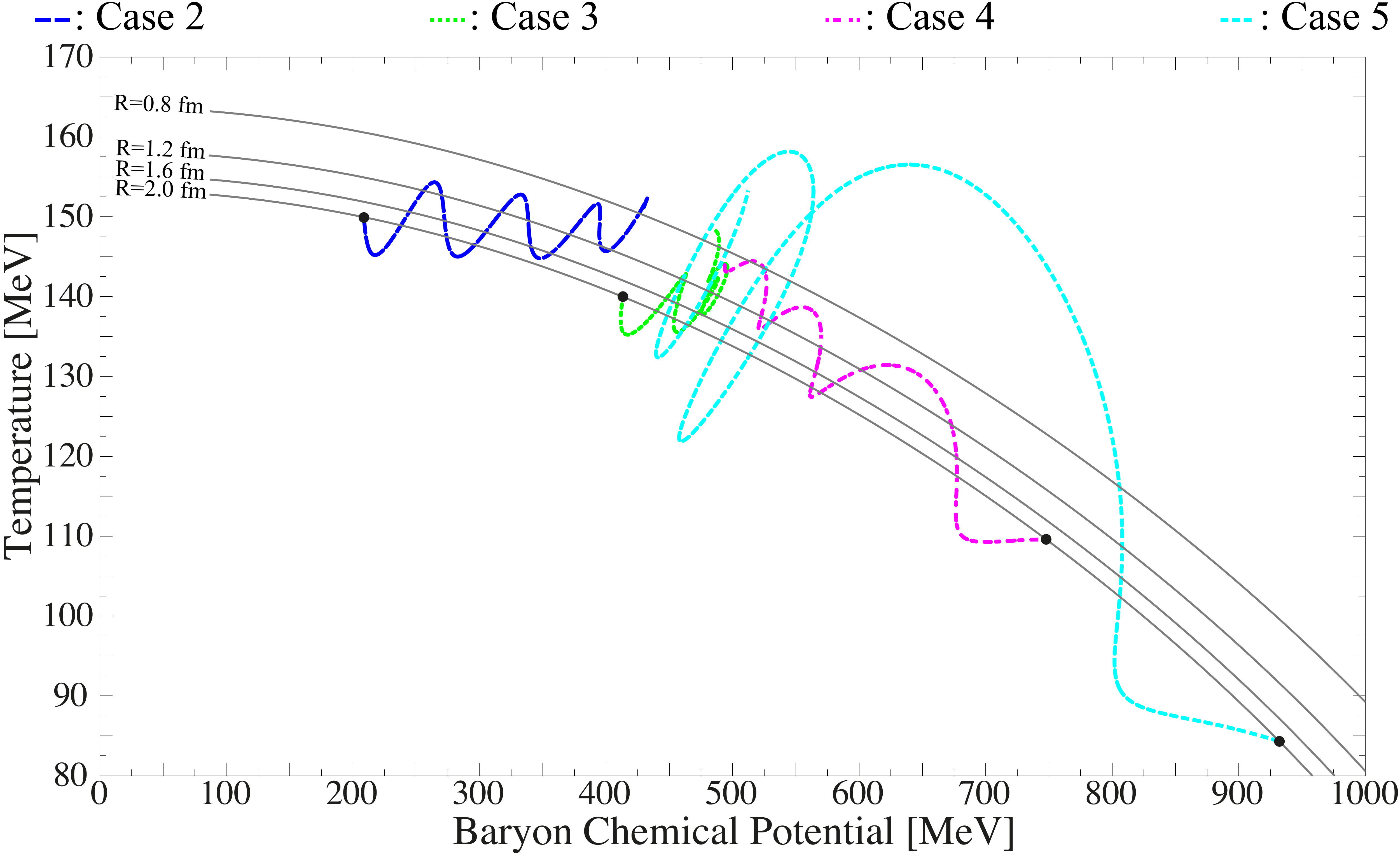}
\caption{\small (color online) Trajectories of the quark droplets in the $\mu_{\rm B} - T$ plane are shown for four different initial conditions (as indicated in the figure). A cut-off at $R = 0.8 \rm \ fm$ is applied. The dots indicate the initial points of the droplet evolution. The four solid, grey lines represent isobars of the net pressure (see Eq. (\ref{PressureBalance})) for four different values of the radius (indicated in the figure) at zero surface velocity.}\label{plot:baryonChemicalPotVStemperature}

\centering
\includegraphics[width=\linewidth]{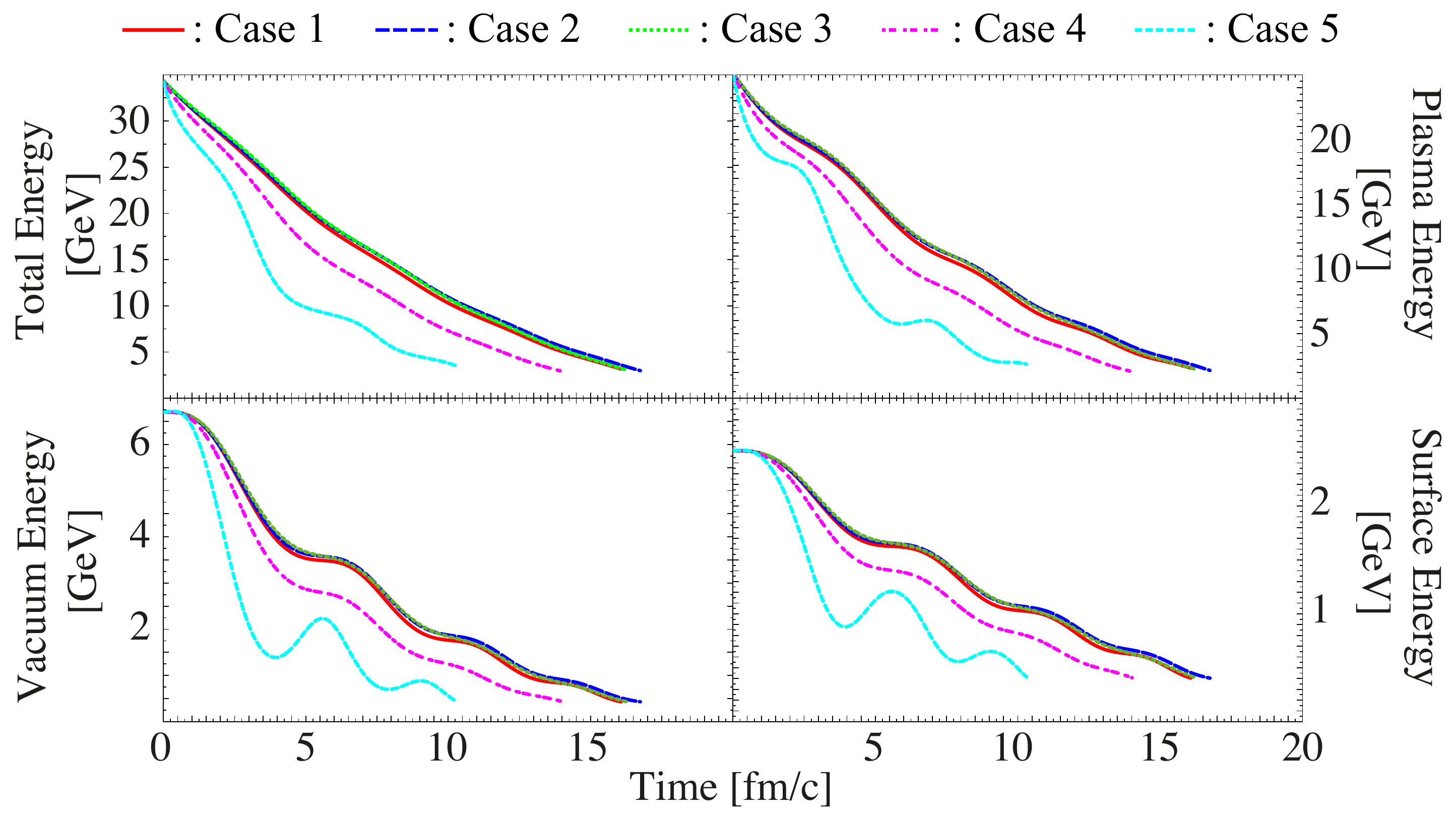}
\caption{\small (color online) The total energy of the droplet together with plasma energy, vacuum energy and surface energy are shown as functions of time for four different initial conditions (indicated in the figure). A cutoff of $R$=0.8 fm is applied.}\label{plot:energy}
\end{figure}

The most remarkable feature of Fig. \ref{plot:evolution} is the convergence of both the baryon density and temperature to some universal values $n_{\rm B} \approx 0.45 \ \rm fm^{-3}$ and $T\approx 150 \ \rm MeV$, especially clear for cases 1-4. Although we start at four very different initial states, both the temperature and the baryon density at the final stage lie quite close for all four cases. The same behaviour is seen in Fig. \ref{plot:baryonChemicalPotVStemperature}, depicting the temperature as function of baryon chemical potential of the droplet for cases 2 to 5, together with four solid grey lines each representing a set of $\mu_{\rm B}$ and $T$ values for four different values of the radius (indicated in the figure) where the thermal pressure outbalances the sum of vacuum pressure and the Laplace pressure i.e. $P_{\rm net}(R,T,\mu_{\rm B})=0$ (see Eq. (\ref{PressureBalance})). This figure shows more clearly, the convergence toward the values $T\approx 150 \ \rm MeV$ and $\mu_{\rm B}\approx 450 \ \rm MeV$. This convergence is linked to emission ratio of baryons versus mesons. Crudely expressed, when a baryon is emitted, the temperature and the baryon chemical potential are reduced. This leads to a lower probability of emitting a baryon afterwards and is thereby enhancing the emission of mesons versus baryons. However, if a meson is emitted only the temperature is changed, while the baryon chemical potential stays constant. Meson emission thereby enhances baryon emission and a balance is established between the temperature and the baryon chemical potential leading to the convergence seen in the dynamical plots. A similar explanation was given in \cite{Bar90}, where the authors have found a constant temperature behaviour during the evolution of their droplets. They explained this constant temperature as a cancellation of cooling due to meson emission and heating due to baryon emission. In our case, however, the general trend is an increasing temperature (see Fig. \ref{plot:evolution}). This is seen best in cases 4 and 5, showing a strong heating of droplets by about 35 MeV and 70 MeV, respectively. Only in case 2, we find a somewhat constant temperature.

Normally, one would associate an evaporation with a cooling of the remaining liquid, as seen in evaporation from a cup of hot water. In our case we have, however, a dynamical feedback from the surface tension: As the droplet decreases in size the pressure from the surface tension is increasing. To counteract the increasing pressure, the thermal pressure inside the droplet can be increased in two ways either by increasing the baryon chemical potential or the temperature of the quarks. However, any enhancement of the baryon chemical potential leads to a higher probability of emitting baryons which again lowers the baryon chemical potential and temperature. Therefore, the only way for the droplet to restore the pressure balance is to increase the temperature of the plasma by creation of quark-antiquark pairs, i.e. mesons.

The time evolution of different contributions to the droplet energy is shown in Fig. \ref{plot:energy}. One can see for all five cases, that the energy drops by about 31 GeV from initial 34 GeV to 3 GeV during the droplet evolution. In the more baryon abundant emission, cases 4 and 5, we find a stronger oscillation of the energy with significant drop of energy during the first 5 fm/$c$. Compared to case 1, the life time is slightly increased in cases 2 and 3, while it is decreased by approximately 3 fm/$c$ and 6 fm/$c$ in cases 4 and 5, respectively. We can conclude from this figure, that the lifetime and the radial evolution of the droplet change significantly when we consider initially very baryon-rich, low-temperature droplets, like in cases 4 and 5. A shorter lifetime is in better agreement with hadronization times extracted from HBT measurements which are about 10 fm/$c$ at LHC and 7 fm/$c$ at RHIC \cite{Aam11}.

\begin{figure}
\centering
\includegraphics[width=\linewidth]{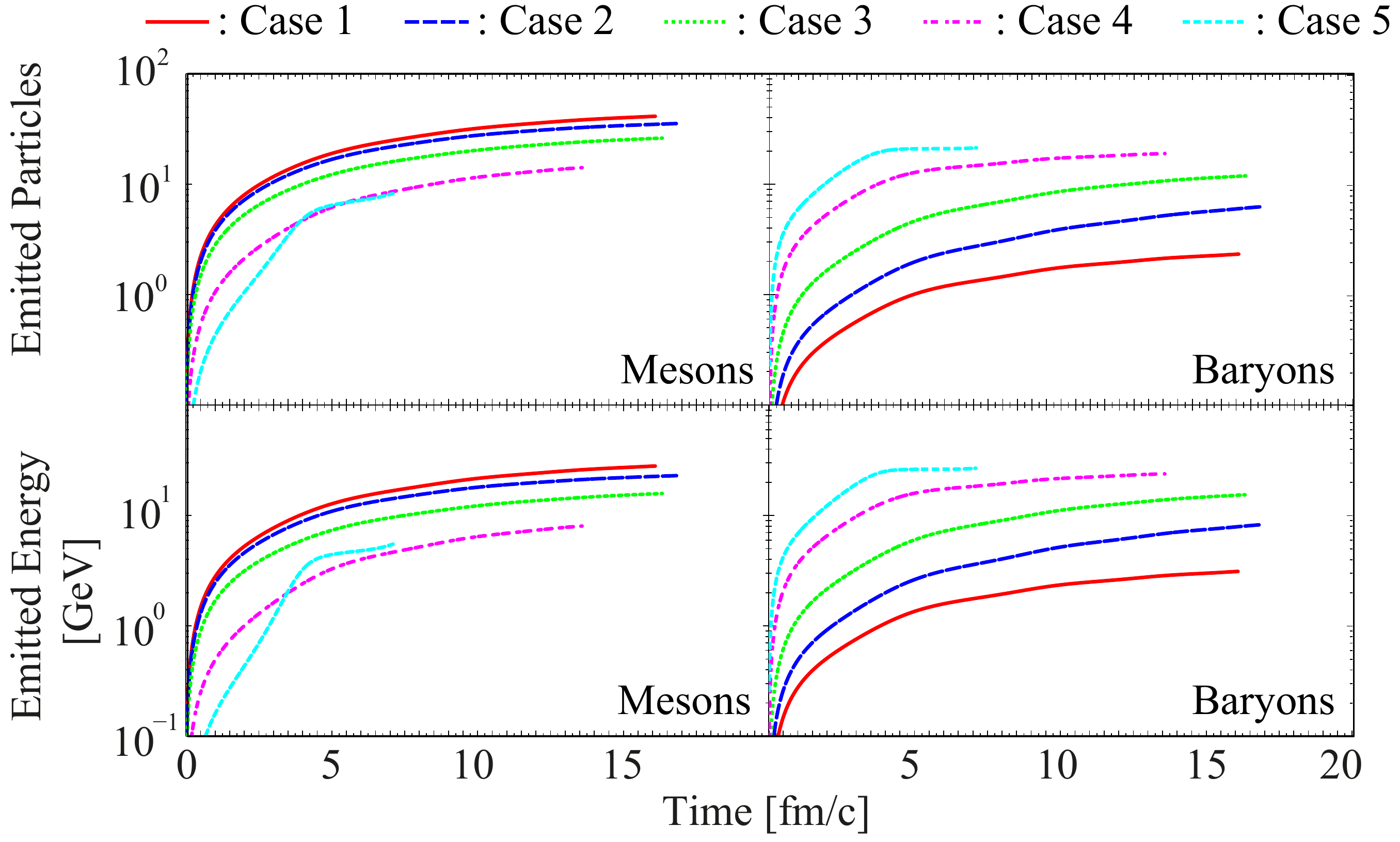}
\caption{\small Total number and energy of both mesons $(\pi, K, \rho, \omega)$ and baryons $(N, \Delta, \Lambda, \Sigma)$ emitted as a function of time for the 5 different cases discussed above. A cut-off at $R = 0.8 \rm \ fm$ is applied.}\label{plot:emission}
\end{figure}

The difference of the lifetimes can be understood from from Fig. \ref{plot:emission}, where the total numbers of emitted mesons and baryon are shown for all 5 cases together with their contribution to the emitted energy. From cases 1 to 5, we see a shift from a very meson abundant emission to a more baryon abundant emission. As the baryons, because of the larger mass, carry away significantly more energy compared to the mesons, this has great influence on the energy balance. The very small increase of the lifetime in case 2 and 3 is an effect of the increased emission of baryons and at the same time reduced emission of anti-baryons and pions due to the initially lower temperature and higher baryon chemical potential. For cases 4 and 5 the baryon emission is larger than the meson emission, which causes a faster drop of energy compared to cases 1 and 2, as seen in Fig. \ref{plot:energy}.

\section{Conclusion}\label{sec:Conclusion}
We have presented a simple hydro-kinetic model of a quark droplet evolution, which takes into account bulk and surface contributions to the energy, collective expansion, hadron emission from the surface and a non-zero net baryon number. We have found, at small values of baryon chemical potential $(\mu_{\rm B} \lesssim T)$, the lifetime and radial behaviour of the droplets do not change significantly as compared with close-to-equilibrium results presented in (I). Only when we consider an initially very baryon-rich, low-temperature system is the lifetime decreased significantly and the radial behaviour of droplets showed similarities with the behaviour found in (I) for the out-of-equilibrium initial conditions.

We have, furthermore, found a convergence of both the temperature and baryon chemical potential toward the values $T \approx 150$ MeV and $\mu_{\rm B} \approx$ 450 MeV. This convergence seems to be linked to the competition in emission of baryons and mesons. A meson emission enhances baryon emission and a balance is established between the temperature and the baryon chemical potential leading to the convergence of dynamical trajectories. Unfortunately, we did not find any tendency toward the formation of cold quark droplets in their evolution.

In the future, we plan to introduce also, the strangeness degree of freedom to study the possibility of strangelet formation.

\section*{Acknowledgements}
The authors thank L.M. Satarov and S. Schramm for fruitful discussions. J.J.B.-B. acknowledges the fellowship received from the Helmholtz International Center for FAIR within the framework of the LOEWE program of the state of Hessen, Germany. I.N.M. acknowledges support provided grant NS-215.2012.2 (Russia).
\newpage 
\bibliography{References}{}
\bibliographystyle{unsrt}

\end{document}